\documentstyle[preprint,revtex]{aps}
\begin{document}
\draft
\hfill
\vbox
{\hbox{\bf NUHEP-TH-92-29}\hbox{\bf DOE-309-CPP-47}\hbox{January 1993}}\par
\thispagestyle{empty}
\begin{title}
{\bf Detection of the heavy Higgs boson at $\gamma\gamma$ colliders}
\end{title}
\author{David Bowser-Chao$^{*}$ and Kingman Cheung$^{**}$}
\begin{instit}
* Center for Particle Physics, University of Texas at Austin, Austin, Texas
78712, USA\\
** Dept. of Physics \& Astronomy, Northwestern University, Evanston,Illinois
60208, USA\\
\end{instit}
\begin{abstract}
\nonum
\section{Abstract}
We consider the possibility of detecting a heavy Higgs boson
($m_H>2m_Z$) in proposed $\gamma\gamma$ colliders through the
semi-leptonic mode $\gamma\gamma \rightarrow H \rightarrow
ZZ \rightarrow q\bar q \ell^+\ell^-$.
 We show that due to the non-monochromatic nature
of the photon beams produced by the laser-backscattering method,  the
resultant  cross section for Higgs production
is  much smaller than the on-resonance cross section and generally
{\it decreases} with increasing collider energy. Although continuum
$ZZ$ production is expected to be negligible, we demonstrate the
presence of and calculate sizeable backgrounds from
 $\gamma\gamma\rightarrow \ell^+\ell^-Z,\,q\bar qZ$, with
$Z\rightarrow q\bar q,\,\ell^+\ell^-$, respectively, and
 $\gamma\gamma\rightarrow t\bar t\rightarrow b\bar
b\ell^+\ell^-\nu\bar\nu$.   This channel may be used to detect a Higgs
of mass $m_H$ up to around 350~GeV at a 0.5~TeV $e^+e^-$ collider,
assuming a nominal yearly luminosity of 10--20~fb$^{-1}$.
\end{abstract}

\newpage
\section{INTRODUCTION}
\label{intro}
The next generation $pp$, $e^+e^-$, and $ep$ colliders will search for the
Higgs boson \cite{hunter} over a wide range of Higgs masses $m_H$.
Detection of a heavy  Higgs ($m_H > 2m_Z$) is considered feasible at
the Superconducting Super Collider (SSC) or CERN Large Hadron Collider
(LHC) through the gold-plated channel, $H\to Z Z\to \ell^+\ell^-
\ell^+\ell^-$\cite{ZZ}, and also at the Next Linear $e^+e^-$ Colliders
(NLC) through the decay of $H\rightarrow WW,\,ZZ\rightarrow
(jj)(jj)$\cite{ee}. Beyond the discovery of the Higgs boson and the
determination of its mass, the measurement of its coupling to photons
is desirable for the information afforded on the Yukawa coupling of
the Higgs to the top quark \cite{BOOS} and the existence of ultraheavy
new particles \cite{chano} . For a heavy Higgs, however, the branching
ratio for $H\to\gamma\gamma$ is too small for accurate experimental
measurement of the $H\gamma\gamma$ coupling at  the SSC, LHC or  NLC.

Proposed $\gamma\gamma$ colliders\cite{teln,beams}, based on the underlying
next-generation linear $e^+e^-$ colliders, provide both a means of detecting a
heavy Higgs \cite{bord} as well as obtaining information on the
$H\gamma\gamma$ coupling, through  direct Higgs production via
photon-photon fusion.  The Higgs may be detected through
 $H\to ZZ,\,W^+W^-$ and $t\bar t$. The latter two channels suffer from
immense tree-level continuum backgrounds\cite{WWback,halzen}, and the
top-quark has not yet been found, so we concentrate on the $H\rightarrow ZZ$
decay channel.
Previous studies of Higgs resonance production
\cite{BOOS,chano,richard,heavy,gunion} at $\gamma\gamma$ colliders,
however,  have noted the absence of a tree-level background to
$\gamma\gamma\to H\to ZZ$.  The signal is itself a
one-loop process, but should be of order $\alpha$ less-suppressed
than the production of $ZZ$ through  box diagrams.  Furthermore, the signal
is almost entirely restricted to the interval  $m_H\pm \Gamma_H$ in
the invariant mass spectrum of $m(ZZ)$ because the Higgs width is very narrow
for $m_H\alt 400$~GeV; for the visible decay channels, we can further
reduce this background by requiring $m(ZZ)$  to be around the Higgs peak.
If the continuum production represented the sole background, detection of the
Higgs via $H\to ZZ$ would depend only on the total event rate. We
shall demonstrate, however, the presence of additional
backgrounds, which  come into play when detection of
the gauge boson pair is taken into account. The purpose of this
paper is to investigate the feasibility of detecting $H\to ZZ$ in light of
these backgrounds, for the various decay modes of the gauge bosons.

As shown below, the non-monochromatic nature of the photon beams
drastically reduces the Higgs production cross section from its on-resonance
value. The ``gold-plated'' detection mode, where $H\to ZZ \to
\ell^+\ell^-\ell^+\ell^-$, is thus marginal for all but very high collider
luminosities because of the small branching fractions incurred. On the other
hand, previous authors \cite{chano,gunion} have pointed out that the most
abundant decay channel, where both $Z$ bosons decay hadronically, is
obscured by the huge continuum background from $\gamma\gamma\to WW$
\cite{WWback}, where the $W$ bosons decay hadronically. The leptonic decay
with neutrinos, where $H\to ZZ \to \ell^+\ell^- \nu\bar\nu$, has been
considered \cite{chano,heavy,gunion}, but we note that the leptonic decay
of $\gamma\gamma\to W^+W^-\to\ell^+\nu\ell^-\bar\nu$ presents an overwhelming
background, even after requiring the lepton pair to reconstruct a $Z$
boson. In the same vein, the decay of $WW\to q\bar{q}\ell\nu$ presents a
large, detector-dependent background to the signature of $\gamma\gamma\to H
\to ZZ \to q\bar q\nu\bar\nu $. Before any cuts, this background turns out
to be about four orders of magnitude larger than the signal \cite{WWback}.
The viability of this channel crucially depends on the efficiency of a cut
on $m(q\bar q)$. A less important, though helpful, measure would be to
additionally veto events with an isolated charged lepton. This mode
furthermore precludes direct reconstruction of the Higgs mass.

This paper thus considers the feasibility of observing the remaining
decay channel:

\begin{equation}
\gamma\gamma \to H \to ZZ \to \ell^+\ell^-q\bar q\,, \label{qqll}
\end{equation}
whereas limits in the
reconstruction of either gauge boson lead to
backgrounds from:
\begin{eqnarray}
\gamma\gamma &\to & q\bar qZ \to q\bar q \ell^+\ell^-\,, \label{qqZ}\\
\gamma\gamma &\to &\ell^+\ell^- Z \to \ell^+\ell^- q\bar q\,,\label{llZ}\\
\gamma\gamma &\to &t\bar t \to b\bar b \ell^+\ell^-
\nu_\ell\bar\nu_\ell\,,\label{ttbar}
\end{eqnarray}
where the charged lepton and quark pairs are required to reconstruct the
$Z$ mass to a given resolution (we note that
process~(\ref{ttbar}) is another potential background to $H\to ZZ\to
q\bar q\nu\bar\nu$).  We will show that with selective acceptance cuts on the
charged leptons and quarks, we can reduce these backgrounds to a manageable
level.

The organization of this paper is as follows: we briefly describe the
laser back-scattering method and the corresponding photon luminosity in
Sec.~\ref{II}, following which Sec.~\ref{III} details the calculation
of the signal and  background processes. Finally, we discuss our
results in Sec.~\ref{IV}, and summarize the conclusions of our analysis in
Sec.~\ref{V}.

\section{Laser Back-scattering Method}
\label{II}

Hard $\gamma\gamma$ collisions at $e^+e^-$ machines can be produced by
directing a low energy (a few $eV$) laser beam at a very small angle
$\alpha_0$, almost head to head, to the incident electron beam.
Through Compton scattering, there are abundant, hard back-scattered
photons in the same direction as the incident electron, which carry a
substantial fraction of the energy of the incident electron.
Similarly, another laser beam can be directed onto the positron beam,
and the resulting $\gamma$ beams effectively make a $\gamma\gamma$
collider. Further technical details may be found in
Refs.~\cite{teln}. Another possibility is to use the beamstrahlung
effect \cite{beams} but this method produces photons mainly in the
soft region \cite{teln}, and depends critically on the beam structure
\cite{beams}. For the production of a heavy Higgs we
need a fairly high energy threshold of the two incoming photons.
Therefore we shall limit our calculations to $\gamma\gamma$ collisions
produced by the laser back-scattering method.

\subsection{Photon Luminosity}

We use the energy spectrum of the back-scattered photon given by \cite{teln}
\begin{equation}
\label{lum}
F_{\gamma /e}(x) = \frac{1}{D(\xi)} \left[ 1-x +\frac{1}{1-x}
-\frac{4x}{\xi(1-x)} + \frac{4x^2}{\xi^2 (1-x)^2} \right] \,,
\end{equation}
where
\begin{equation}
\label{D_xi}
D(\xi) = (1-\frac{4}{\xi} -\frac{8}{\xi^2}) \ln(1+\xi) + \frac{1}{2} +
\frac{8}{\xi} - \frac{1}{2(1+\xi)^2}\,,
\end{equation}
$\xi=4E_0\omega_0/m_e^2$, $\omega_0$ is the energy of the incident laser
photon, $x=\omega/E_0$ is the fraction of the incident electron's
energy carried by the back-scattered photon, and the maximum value $x_{\rm
max}$ is given by
\begin{equation}
x_{\rm max}= \frac{\xi}{1+\xi}\,.
\end{equation}
It is seen from Eq.~(\ref{lum}) and (\ref{D_xi}) that the portion of
photons with  maximum energy  grows with $E_0$ and $\omega_0$.
A large $\omega_0$, however, should be avoided so that the
back-scattered photon does not interact with the incident photon and create
unwanted $e^+e^-$ pairs.  The threshold for  $e^+e^-$ pair creation is
$\omega \omega_0 > m_e^2$, so we require $\omega_{\rm max}
\omega_0 \alt m_e^2$.  Solving $\omega_{\rm max}\omega_0=m_e^2$, we
find
\begin{equation}
\xi = 2(1+\sqrt{2}) \simeq 4.8 \,.
\end{equation}
For the choice $\xi=4.8$ one finds $x_{\rm max}\simeq 0.83$,
$D(\xi)\simeq 1.8$, and
$\omega_0=1.25(0.63)$~eV for a 0.5(1) TeV $e^+e^-$ collider.
Here we have assumed that the electron, positron and back-scattered
photon beams are unpolarized.  We also assume that,
on average, the number of back-scattered photons produced per electron
is 1 (i.e., the conversion coefficient $k$ equals 1).

\section{Signal \& Background Calculation}
\label{III}
\label{sec:III}

In calculating the results presented below, we folded the photon
distribution function into the hard scattering cross section $\hat \sigma$ for
each subprocess. The total cross section $\sigma$ is given by
\begin{equation}
\sigma(s) = \int_{\tau_{\rm min}}^{x_{\rm max}^2}  d\tau \label{totsigma}
\int_{\tau/x_{\rm max}}^{x_{\rm max}} {dx_1 \over x_1}
F_{\gamma/e}(x_1) F_{\gamma/e}(\tau/x_1) \hat \sigma (\hat s = \tau s)\,,
\end{equation}
where
\begin{equation}
\tau_{\rm min}   =   \frac{(M_{\rm final})^2}{s}\,,
\end{equation}
and $M_{\rm final}$ is the sum of masses  of the final state particles.

\subsection{$\gamma\gamma\to H\to ZZ$}
\label{subsec:sigcalc}

The contributing Feynman diagrams are depicted in Fig.~\ref{feyn1} (the
subsequent decay of the gauge bosons, $ZZ\to\ell^+\ell^-q\bar q$, is not
shown). All massive charged particles contribute to the loop \cite{hunter};
in particular, this process could be used to probe the Higgs sector with  a
heavy top quark \cite{BOOS}, or to detect the presence of new ultra-heavy
fermions \cite{chano} or non-SM $W'$ charged gauge bosons. In this paper,
we assume only SM particles in the triangular loop; in particular, we
have used a top-quark mass of 150~GeV.

The hard scattering cross section for $\gamma\gamma \to H\to Z Z$, for
monochromatic photons with $m(\gamma\gamma)=\sqrt{\hat s}$, is given by:
\begin{equation}
\label{sigmahat}
\hat\sigma(\hat s) =
 {{8\pi \Gamma(H^*\to\gamma\gamma)\Gamma(H^*\to Z Z)}
	\over
  {(\hat s - m_H^2)^2 + \Gamma_H^2 m_H^2}}\label{hard}\,,
\end{equation}
where the partial widths in the numerator are those of a virtual Higgs of
mass $m_{H^*}=\sqrt{\hat s}$. The on-resonance cross section $\sigma_0$ is
simply given by $\sigma_0 = \hat\sigma(m_H^2)$. We have calculated the
total cross section $\sigma(\gamma\gamma\to H\to ZZ)$ by folding in
Eq.~(\ref{hard}) with the photon luminosity functions as in
Eq.~(\ref{totsigma}).

It is instructive to apply the narrow width approximation to
this cross section. For the range of $m_H$ considered here ($m_H \alt
400$~GeV), the true width $\Gamma_H$ is quite small. Then the partial widths
can be approximated by their on-shell values, obtaining:
\begin{equation}
\label{sigma}
\sigma(s) = \sigma_0 \,\,\, \left(\frac{\pi m_H \Gamma_H}{s}\right)
         \int^{x_{\rm max}}_{m_H^2/(x_{\rm max}s)}
         F_{\gamma/e}(x_1) F_{\gamma/e}(\tau/x_1) \frac{dx_1}{x_1}\,,
\end{equation}
where $\tau=m_H^2/s$ and $\sqrt{s}$ is the center-of-mass energy of the
parent $e^+e^-$ collider. From Eq.~(\ref{sigma})  we anticipate that the
total cross section should roughly decrease with increasing collider
energy $\sqrt{s}$. However, as the maximum energy of the
photon-photon collider ($x_{\rm max} \sqrt{s}$) just passes $m_H$,
$\sigma$ {\it increases} with $\sqrt{s}$ because of enhancement by the
factor $\log(x_{\rm max}^2s/m_H^2)$ from the integration of $x_1$.

The photon is known to have anomalous quark and gluon contents
\cite{witten}. The Higgs can thus also be produced by the gluon-gluon
fusion, where the contributing Feynman diagrams are similar to those of
photon-photon fusion, but with only quarks in the triangular loop. Since we
are searching for a heavy Higgs ($m_H >2m_Z$), the threshold value for $\sqrt
{\tau}$
(equal to $2m_Z/\sqrt{s}$) is about 0.1 and 0.4 at $\sqrt{s}= 1.5$ and
0.5~TeV, respectively. For this range of $\sqrt{\tau}$,
the effective
gluon-gluon luminosity is extremely small \cite{halzen}, so that the
contribution from gluon-gluon fusion is negligible.
Furthermore, gluon-photon fusion does not contribute at one-loop level
because the Higgs is color neutral.

To obtain the total cross section for the decay channel $\gamma\gamma\to
H\to ZZ\to \ell^+\ell^-q\bar q$, we simply multiply by the relevant branching
fractions. To properly assess the effect of cuts on this mode, however, we
included full spin-correlation of the $Z$-boson decay products. We thus
alternately replaced the width $\Gamma(H^*\to Z Z)$ in Eq.~(\ref{hard})
with  $\Gamma(H^*\to Z_L Z_L)$ and $\Gamma(H^*\to Z_T Z_T)$, and
generated the respective angular distribution of the decay products.

\subsection{$\gamma\gamma\to q\bar q Z$ and $\ell^+\ell^- Z$}

The contributing Feynman diagrams for the process $\gamma\gamma\to
q\bar qZ$ are shown in Fig.~\ref{feyn2}(a).  Those for $\gamma\gamma\to
\ell^+\ell^- Z$ are obtained with the replacement of $q$ with $\ell$.
Here again we have not depicted the  subsequent decay of the gauge
bosons into either a quark or charged-lepton  pair. The expressions for
 Feynman amplitudes for the  similar process $\gamma\gamma\to t\bar t
Z$ have been presented in Ref.~\cite{cheung}. Those for $q\bar qZ$ and
$\ell^+\ell^- Z$ are easily obtained by replacing the top-quark with
the lighter quarks ($u,d,s,c,b$) and charged leptons ($e,
\mu$), and substituting with the corresponding $g_{Zff}$ and $g_{\gamma
ff}$ couplings. In our calculations of these backgrounds, we have set
all quark and lepton masses to $m_f=m_q=m_l=0.1$~GeV. In the next
section, we will show that with our acceptance cuts the cross sections of
these backgrounds are independent of
the choice of $m_f$. We have also assumed that we can
distinguish a tau from a muon, electron, or quark.

The processes of Eqs.~(\ref{qqZ}) and (\ref{llZ}) have the same final state
particles
$q\bar q \ell^+\ell^-$ as the signal. These backgrounds mimic the signal
when the invariant-mass of the continuum $q\bar q$ or $\ell^+\ell^-$ pair
is within detector resolution of the $Z$ mass. We use a rather conservative
mass constraint
\begin{equation}
\label{mass}
 |m(q\bar q\,{\rm or}\,\ell^+\ell^-) - m_Z| < 8\;{\rm GeV}\,,
\end{equation}
to reconstruct the ``fake $Z$''.
We have included full spin correlation for the subsequent
decay of the real $Z$ into either $q\bar q$ or $\ell^+\ell^-$ pair.

In contrast to the signal, we must consider the photon-gluon contribution
to $q\bar qZ$ production. However, the gluon distribution function inside
photon has large uncertainties because of limited experimental data.  We
have employed two different parameterizations,
Levy-Abramowicz-Charchula set III (LAC3)\cite{LAC3} and Drees-Grassie
(DG)\cite{DG}, for the photon structure functions. We have set the
scale $Q^2=\hat s/4$ for both the photon structure functions and
$\alpha_s$ (evaluated to second order), and $\Lambda_4$ to be
0.2 and 0.4~GeV for the LAC3 and DG sets, respectively.

\subsection{$\gamma\gamma\to t\bar t$}

The contributing Feynman diagrams are shown in Fig.~\ref{feyn2}(b).
The top-quark pair  decays into $b\bar bWW$; if the $W$
bosons decay leptonically into $\ell^+\ell^-\nu_\ell\bar\nu_\ell (\ell=e,\mu)$,
 this process could be mistaken as signal if both $m(q\bar q)$ and
$m(\ell^+\ell^-)$ are in the vicinity of the $Z$ mass. We have chosen
$m_t=150$~GeV for illustrative purposes. Even with a possible doubling
due to smaller top-quark  mass or enhancement from threshold effects
\cite{halzen},  this process turns out to be negligible due to the
additional branching fractions for the $W$ decays and the simultaneous
$Z$-mass constraints of Eq.~(\ref{mass})  on both $m(q\bar q)$ and
$m(\ell^+\ell^-)$. We have included full spin correlations in the
decays of $t\bar t$ and the subsequent decays of  $W^+W^-$.

\section{RESULTS \& DISCUSSION}

\label{IV}

The dependence of the cross section $\hat\sigma$ for
$\gamma\gamma \rightarrow H\rightarrow ZZ$ on the center-of-mass energies
$\sqrt{s_{\gamma\gamma}}$ of the incoming photons is shown in
Fig.~\ref{cross}(a) for a range of Higgs masses $m_H$ from 200--500~GeV,
assuming monochromatic photons;  this figure agrees with the results in
Ref.~\cite{BOOS}.
As discussed above, however, laser back-scattering produces
photons with a spread in energies, so that we must  fold $\sigma_0$ in
with the effective photon-photon luminosities to obtain the actual
cross section.  We show the dependence of  the total cross section
$\sigma(e^+e^-\to\gamma\gamma \rightarrow H\rightarrow ZZ)$ on the
center-of-mass energies $\sqrt{s_{e^+e^-}}$ of the parent $e^+e^-$
collider for the same range of $m_H$ in Fig.~\ref{cross}(b).
As previously discussed, the resultant cross sections are
greatly reduced from their on-resonance values $\sigma_0$, and generally fall
with $\sqrt{s_{e^+e^-}}$. The cross sections, however, increase in the
 small range of $\sqrt{s_{e^+e^-}} \agt m_H$,  which is most
apparent for $m_H\agt 400$~GeV.
We also show the dependence of the actual cross section on the Higgs-boson
mass $m_H$ for $\sqrt{s_{e^+e^-}}=0.5$ and 1~TeV in Fig.~\ref{cross}(c).
We can see that the maximum cross section occurs at slightly above
$m_H=200$~GeV, then falls quite sharply with further increase in $m_H$.
It should be noted that the
smallness of these cross sections precludes the use of the gold-plated
mode $H\to ZZ\to\ell^+\ell^-\ell^+\ell^-$, for which the optimal case
of $m_H=200$~GeV at $\sqrt{s_{e^+e^-}}= (0.3)\,0.5$~TeV and an integrated
luminosity of 100~fb$^{-1}$ would yield only about (13) 10 events.
Furthermore, the drop in direct Higgs production with
increasing collider energies suggests that associated production of the Higgs
with a $t\bar t$ pair\cite{BOOS,cheung} or $W^+W^-$ pair \cite{assoc}
would be more  appropriate for $\sqrt{s_{e^+e^-}} \agt 1.5$~TeV.

With only the mass constraints of Eq.~(\ref{mass}), the $t\bar t$
background of Eq.~(\ref{ttbar}), as  remarked above, turns out to be
very small, totalling about $\sim 0.08$~fb for $m_t=150$~GeV. The
major backgrounds come from  the processes of Eqs.~(\ref{qqZ}) and (\ref{llZ}),
which total to about  30--90~fb for the fermion masses $m_f=$~100~MeV to
0.5~MeV. The cross section for $\gamma\gamma\to\ell^+\ell^- Z$ is
approximately ten times larger than $\gamma\gamma\to q\bar q Z$
because of the difference between the branching fractions of $Z$ into
quarks and charged leptons, respectively, even after including the
contribution to the latter process from photon-gluon fusion.
We expect, due to our
conservative cuts imposed on the lepton pair invariant mass, that
these backgrounds could in practice be smaller (in accordance with
actual detector  resolution for $m(\ell^+\ell^-)$).

Because the production of $f\bar fZ$, where $f=\ell,q$, would be
collinearly divergent in the beam direction
for $m_f$ equal to zero, we expect a
cut on the angles between the  decay products and the beam direction
to be efficacious; some
cut of this nature would of course be required due to limited detector
coverage. This  expectation is clearly borne out by Fig.~\ref{cos},
which shows the dependence  of the integrated cross sections
$\sigma(z<z_{\rm cut})$ on $z_{\rm cut}$,  where $z$ is defined by
\begin{equation}
\label{zdef}
 z = \max\left\{|\cos\theta_i|\right\} \,, \qquad i=q,\bar q, \ell^+ \ell^-
\end{equation}

It can be seen that the background is rather sharply-peaked in the forward
and backward regions. We therefore require
\begin{equation}
\label{coscut}
z_{\rm cut}=0.95
\end{equation}
which corresponds to requiring an angle of about $8^\circ$ from the beam-pipe.
This cut has the effect of drastically reducing the backgrounds of
Eq.~(\ref{qqZ}) and (\ref{llZ}) but only mildly affecting the signal
and the $t\bar t$ background of Eq.~({\ref{ttbar}).  As described
above, we have taken the fermion-masses $m_f=0.1$~GeV in presenting
the background calculation of Eq.~(\ref{qqZ}) and (\ref{llZ}). We have
confirmed that the results with the above cut on $z$ are independent
of the fermion-mass $m_f$ for $m_f\alt 1$~GeV.

Because all the decay products can be detected after requiring
separation from the beam-pipe by the cut on $z$ above, we can reconstruct the
total invariant mass of the quark and charged-lepton pairs.
In Fig.~\ref{m(zz)} we show the
dependence of the differential cross section $d\sigma/dm(ZZ)$ on the invariant
mass of the $ZZ$ pair, after the acceptance cuts of $z<0.95$
and the mass constraint of Eq.~(\ref{mass}).  The Higgs-boson
signal forms a sharp peak for $m_H \alt 300$~GeV and a less sharp but still
distinguishable peak for $m_H=350$ and 400~GeV. We define
the signal as the cross section under the Higgs peak in the interval
\begin{equation}
\label{mzzcut}
m_H\pm \Gamma\,,\qquad  {\rm where}\; \Gamma={\rm max}(\Gamma_{\rm
resolution},\, \Gamma_H)\,,
\end{equation}
and we take a rather conservative $\Gamma_{\rm resolution}=10$~GeV.
The same cut is applied to each background process.
With better resolution in $m(ZZ)$, of course, the background can be
further reduced.

Cross sections for the signal and backgrounds for
various Higgs masses and collider energies are presented in Table~\ref{table1}.
We assume that the Higgs-boson  signal can be discovered if there are six
or more events under the peak and the significance $S/\sqrt{B}$ of this
signal is greater than four.  The significances for integrated luminosities of
10, 20, 50 and 100~fb$^{-1}$ are presented in Table~\ref{table2}.
It is clear that
luminosity, rather than collider energy $\sqrt{s_{e^+e^-}}$, is the
key to detecting a heavy Higgs through photon-photon fusion production.
Furthermore, detection of lower Higgs masses ($m_H \alt 350$~GeV) is
easier in this mode at {\it lower} collider energies. With a
luminosity of 10--20~fb$^{-1}$, a $\sqrt{s_{e^+e^-}}=0.5$~TeV
collider could detect Higgs bosons for $m_H$ up to around 350~GeV. A
higher collider energy would necessitate a correspondingly higher
luminosity. A luminosity of 100~fb$^{-1}$ would be required to probe
Higgs masses up to 400~GeV.

\section{Conclusion}}
\label{V}

We have shown that direct production of a heavy Higgs at
$\gamma\gamma$ colliders offers a feasible discovery mode, provided
that the energy of the underlying $e^+e^-$ collider is not too much
greater than $m_H$. Due to the backgrounds  principally
arising from the immense production of $W^+W^-$ pairs, the optimal
detection channel is $H\to ZZ\to q \bar q \ell^+\ell^-$. Although the
continuum background $\gamma\gamma\to ZZ$ to this process is minimal,
large backgrounds arise from the production of
$\gamma\gamma\to q\bar q Z,\, \ell^+\ell^- Z$.
An angular cut requiring  the quark and charged-lepton pairs to be
away from the beam-pipe can drastically reduce these backgrounds.
With this acceptance cut and the $Z$-mass constraints on the quark and
charged-lepton pairs, we should be able to discover a heavy Higgs-boson signal
for $m_H$ up to 300~GeV at a $\sqrt{s_{e^+e^-}}=0.5$~TeV
$e^+e^-$ collider, assuming  a nominal yearly
luminosity of 10~fb$^{-1}$, and up to 350~GeV for a luminosity of
20~fb$^{-1}$.  The upper limit of applicablity of this mode is $m_H
\alt 400$~GeV, for which one requires a luminosity of the order 100~fb$^{-1}$.

\acknowledgements
We are grateful for helpful discussions with D.~A.~Dicus, C.~Kao, and
A.~L.~Stange. This work was supported by the U.~S. Department of
Energy, Division of High Energy Physics, under Grants
DOE-FG02-91-ER40684 (K.~C.) and DOE-FG05-85ER40200 (D.~B.-C.).
Computing resources were provided in part by the University of Texas
Center for High Performance Computing.


\newpage
\figure{\label{feyn1}
Contributing Feynman diagrams for the signal process
$\gamma\gamma\to H \to ZZ$.  The cross diagram of the incoming  photons
is not shown.}

\figure{\label{feyn2}
Contributing Feynman diagrams for the background processes
(a) $\gamma\gamma \to q \bar q (\ell\bar\ell) Z$, and
(b) $\gamma\gamma\to t\bar t$.  Cross diagrams of the incoming  photons
are not shown.}

\figure{\label{cross}
The dependence of the total cross section for the signal $\gamma\gamma\to H\to
ZZ$ on (a) $\sqrt{s_{\gamma\gamma}}$ for monochromatic photons with
$m_H=200,300,400,$ and 500~GeV, (b) $\sqrt{s_{e^+e^-}}$ of the parent $e^+e^-$
collider for the same range of $m_H$, and (c) $m_H$ for
$\sqrt{s_{e^+e^-}}=0.5$ and 1~TeV.}

\figure{\label{cos}
The integrated cross section $\sigma(z<z_{\rm cut})$ versus $z_{\rm cut}$,
where $z = \max\left\{|\cos\theta_i|\right\}\,,
\, i=q,\bar q, \ell^+ \ell^- $, for the signal and the total background
(indicated by the dashed curve) with $m_H=200,\,300,\,400$~GeV
and $m_t=150$~GeV at $\sqrt{s_{e^+e^-}}=0.5$~TeV. The only cut
employed is  $|m(q\bar q\;{\rm or}\; \ell^+\ell^-) - m_Z| < 8$~GeV.
The curves extend to $z_{\rm cut}$=0.999\,.}

\figure{\label{m(zz)}
The differential cross section $d\sigma/dm(ZZ)$ versus the invariant mass
$m(ZZ)$ for the signal with various Higgs masses for
$m_t=150$~GeV at $\sqrt{s_{e^+e^-}}=0.5$~TeV, after the acceptance cut of
 $z <0.95$ and the $Z$-mass constraints of $|m(q\bar q\;{\rm or}\;
\ell^+\ell^-) - m_Z| < 8$~GeV.  The sum of the backgrounds is
indicated by the dashed curve.}

\newpage
\widetext
\begin{table}
\caption{Cross sections in fb for the signal ($\gamma\gamma\rightarrow
H\rightarrow ZZ \rightarrow q\bar q \ell^+\ell^-,\,\ell=e,\,\mu$) and
the  backgrounds, for various values of $m_H$, at
$\sqrt{s}=0.5,\,1$ and 1.5~TeV $e^+e^-$ colliders.
The cross sections  already include the branching fraction to
the $q\bar q\ell^+\ell^-$ final state. The acceptance cuts are
$z<0.95$, $|m(\ell\ell\,{\rm or}\,qq) - m_Z|<8$~GeV, and
$m_H-\Gamma <m(ZZ)<m_H+\Gamma$, where $\Gamma$ is defined in
Eq.~(\ref{mzzcut}).
\label{table1}}
\renewcommand{\arraystretch}{0.8}
\begin{tabular}{|c|c|ccccc|}
$m_H$ &
$\gamma\gamma\to H$ &
$\gamma\gamma\to \ell^+\ell^-Z$ &
$\gamma\gamma\to q\bar q Z$  &
\multicolumn{2}{c}{$\gamma g \to q\bar q Z$}  &
$\gamma\gamma \to t\bar t$ \\
&&&& DG & LAC3 & \\
\hline
& &\multicolumn{4}{c}{(a) $\sqrt{s} = 0.5$ TeV} &\\
$\Gamma=\infty$      & ---  & 1.7   & 0.13  & 0.0043 & 0.016 & 0.064\\
200           & 1.6  & 0.29  & 0.021  & 0.0016 & 0.0055& 0.018\\
250           & 1.4  & 0.22  & 0.016 & 0.00042& 0.0017& 0.0099\\
300           & 0.92 & 0.13  & 0.0093 & 6.8$\times 10^{-5}$ & 0.00028 & 6.7
$\times10^{-4}$ \\
350           & 0.38 & 0.12  & 0.0088 & 1.2$\times 10^{-5}$ & 4.2$\times
10^{-5}$ & 0.0 \\
400           & 0.099& 0.070 & 0.0051 & 6.4$\times 10^{-7}$ & 1.4 $\times
10^{-6}$ & 0.0 \\
\hline
& &\multicolumn{4}{c}{(b) $\sqrt{s} = 1$ TeV} &\\
$\Gamma=\infty$      & ---  & 0.82   & 0.059  & 0.016 & 0.042    & 0.080\\
200           & 0.52 & 0.084  & 0.0061 & 0.0038  & 0.0093 & 0.020\\
250           & 0.53 & 0.084  & 0.0061 & 0.0022  & 0.0055 & 0.012 \\
300           & 0.42 & 0.055  & 0.0040 & 0.00087 & 0.0023 & 0.0027\\
350           & 0.21 & 0.063  & 0.0045 & 0.00056 & 0.0016 & 0.0011\\
400           & 0.11 & 0.071  & 0.0052 & 0.00038 & 0.0011 &
1.2$\times10^{-4}$\\
\hline
& &\multicolumn{4}{c}{(c) $\sqrt{s} = 1.5$ TeV} &\\
$\Gamma=\infty$      & ---  & 0.40   & 0.030  & 0.019   & 0.045  & 0.045 \\
200           & 0.23 & 0.038  & 0.0027 & 0.0036  & 0.0081 & 0.013\\
250           & 0.23 & 0.037  & 0.0027 & 0.0024  & 0.0055 & 0.0068\\
300           & 0.20 & 0.025  & 0.0018 & 0.0012  & 0.0027 & 0.0023\\
350           & 0.11 & 0.032  & 0.0023 & 0.0010  & 0.0024 &
3.7$\times10^{-4}$\\
400           & 0.060& 0.037  & 0.0026 & 0.00086 & 0.0021 & 9.6$\times10^{-5}$
\end{tabular}
\end{table}

\begin{table}
\caption{Cross sections for the signal($S$) and the total
background($B$) in fb, and the significance ($S/\sqrt{B}$) of the signal, for
$m_H=200,\,250,\,300,\,350,\,400$~GeV at $\sqrt{s}=0.5,\,1$ and 1.5~TeV
$e^+e^-$ colliders with various integrated luminosities.
\label{table2}}
\begin{tabular}{|c|cc|cccc|}
$m_H$   &   $S$    &    $B$   & & \multicolumn{2}{c}{$S/\sqrt{B}$} & \\
        &          &          &  10fb$^{-1}$ & 20fb$^{-1}$ & 50fb$^{-1}$ &
100fb$^{-1}$ \\
\hline
&&&\multicolumn{3}{c}{(a) $\sqrt{s}=0.5$ TeV} &\\
200     &   1.6    &    0.34     & 8.7 & 12  & 19 & 27 \\
250     &   1.4    &    0.25     & 8.9 & 12  & 20 & 28     \\
300     &   0.92   &    0.14     & 7.8 & 11    & 17 & 25     \\
350     &   0.38   &    0.13     & 3.3 &  4.7  & 7.4  & 11   \\
400     &   0.099  &    0.075    & 1.1 &  1.6  & 2.6  &  3.6  \\
\hline
&&&\multicolumn{3}{c}{(b) $\sqrt{s}=1$ TeV} &\\
200     &   0.52    &   0.12     & 4.7 &  6.7  & 11 & 15   \\
250     &   0.53    &   0.11     & 5.1 &  7.1  & 11 & 16    \\
300     &   0.42    &   0.064    & 5.3 &  7.4  & 12 & 16   \\
350     &   0.21    &   0.070    & 2.5 &  3.5  &  5.6 &  7.9  \\
400     &   0.11    &   0.077    & 1.3 &  1.8  &  2.8 &  4.0\\
\hline
&&&\multicolumn{3}{c}{(c) $\sqrt{s}=1.5$ TeV} &\\
200     &   0.23    &    0.062   & 2.9 &  4.1  &  6.5 &  9.2   \\
250     &   0.23    &    0.052   & 3.2 &  4.5  &  7.1 &  10   \\
300     &   0.20    &    0.032   & 3.5 &  5.0  &  7.9 &  11 \\
350     &   0.11    &    0.037   & 1.8 &  2.5  &  4.0 &   5.7  \\
400     &   0.060   &    0.042   & 0.9 &  1.3  &  2.1 &  2.9\\
\end{tabular}
\end{table}

\end{document}